\documentclass[journal]{./IEEEtran}

\usepackage{verbatim}
\usepackage{graphicx}
\usepackage[cmex10]{amsmath}
\usepackage{amssymb,amsfonts}
\usepackage{multirow}
\usepackage{theorem}

\newcommand\MYhyperrefoptions
{
	bookmarks=false, % true or false
	bookmarksnumbered=true,
	pdfpagemode={UseOutlines},
	plainpages=false,
	pdfpagelabels=true,
	colorlinks=true,
	linkcolor={black},
	citecolor={black},
	urlcolor={black},
	pdftitle={}, %<!CHANGE!
	pdfcreator={LaTex},
}

\usepackage{cite}
\usepackage[\MYhyperrefoptions,breaklinks=true,]{hyperref}
\usepackage{breakurl}
\usepackage{algorithm}
\usepackage{algorithmic}

\usepackage{bbm}

\usepackage{flushend}  % make final page aligned
\usepackage{caption}
\begin{document}

\title{Deep Deterministic Policy Gradient for Relay Selection and Power Allocation in Cooperative Communication Network}

\author{Yuanzhe~Geng,
	Erwu~Liu,~\IEEEmembership{Senior~Member,~IEEE,}
	Rui~Wang,~\IEEEmembership{Senior~Member,~IEEE,}
	Yiming~Liu,~\IEEEmembership{Graduate~Student~Member,~IEEE,}
	Jie~Wang,
    Gang~Shen,
    and Zhao~Dong
	% <-this % stops a space
	\thanks{Yuanzhe Geng, Erwu Liu, Rui Wang, Yiming Liu, and Jie Wang are with the College of Electronics and Information Engineering, Tongji University, Shanghai 201804, China, E-mail: yuanzhegeng@tongji.edu.cn, erwu.liu@ieee.org, ruiwang@tongji.edu.cn, ymliu\_970131@tongji.edu.cn, 1910688@tongji.edu.cn.}% <-this % stops a space
    \thanks{Gang Shen is with Shanghai Bell Co.Ltd., Shanghai 201206, China, E-mail: gang.a.shen@nokia-sbell.com.}
    \thanks{Zhao Dong is with China Mobile Communications Corp., Beijing 100032, China, E-mail: 13910086047@139.com.}
    % <-this % stops a space
}



\maketitle

\begin{abstract}
    Perfect channel state information (CSI) is usually required when considering relay selection and power allocation in cooperative communication.
    However, it is difficult to get an accurate CSI in practical situations.
    In this letter, we study the outage probability minimizing problem based on optimizing relay selection and transmission power.
	%Cooperative communication is an effective approach to improve spectrum utilization.
	%When considering relay selection and power allocation in cooperative communication, most of the existing studies require the assumption of channel state information (CSI).
	%However, in practical situations, it is usually difficult to get an accurate CSI.
	%In this letter, we study the outage probability minimizing problem in a two-hop cooperative communication scenario, to improve the Quality-of-Service of the system through appropriate relay selection and power allocation.
	We propose a prioritized experience replay aided deep deterministic policy gradient learning framework, which can find an optimal solution by dealing with continuous action space, without any prior knowledge of CSI.
    %The proposed method can deal with continuous action space, which is more advanced than other existing reinforcement learning (RL) based approaches.
	%The key difference from other reinforcement learning (RL) approaches is that, our method can deal with continuous action space.
	Simulation results reveal that our approach outperforms reinforcement learning based methods in existing literatures, and improves the communication success rate by about $4\%$.

\end{abstract}


% Introduction
\section{Introduction}\label{sect_intro}
In recent years, cooperative communication has been paid much attention, for it can help realizing resource collaboration between different nodes and obtaining diversified benefits in multi-user scenario \cite{8275026}.
In cooperative communication, an outage occurs when the received signal-to-noise ratio (SNR) falls below a certain threshold \cite{8435942}, and outage probability is usually used as a metric to measure the Quality-of-Service (QoS) of communication system.
In order to minimize outage probability and improve QoS, it is intuitive to optimize relay selection and power allocation schemes.
Traditional methods usually establish a probabilistic model based on the distribution assumption of channel uncertainty \cite{7438886,7105886,7313006}, and then design relay or power optimizing scheme.
It should be noted that assuming an exact channel state information (CSI) is usually impractical because of the inevitable noise.
Therefore, artificial assumptions about channel state distribution may bring estimation bias and mislead the final decision.
%And these traditional methods can hardly be further applied to other situations.

Reinforcement learning (RL) is one of the three paradigms of machine learning.
RL methods use an agent, which can be taken as an intelligent robot, to interact with and learn from the communication environment, and thus do not need any prior knowledge or assumptions about the environment.
In addition, different from other machine learning methods, RL methods do not require data sets, because all the training data is obtained through continuous interaction \cite{RL:intro2}.
After online data collecting and offline learning, the well-trained network models can be implanted into corresponding equipments for practical use.

To address the aforementioned issue in cooperative communication, several methods have been proposed with the help of RL, by using which optimizing strategy can be directly learned from original communication environment.
Khan \textit{et al.} used SARSA-$\lambda$ algorithm to make an adaptive power allocation \cite{8355169}.
In \cite{6954557} and \cite{9072416}, Q-learning algorithm was employed to help power control and relay selection, respectively.
In \cite{9137340, 8930580, 2011.04891}, the authors developed deep Q network (DQN), which is a combination of RL and deep neural network (DNN), for relay-aided communication.
Further, \cite{8718358} introduced convolutional neural network (CNN) to study relay features in several previous time slots, then the output of CNN is used for value estimation in DQN.
Although such modification obtained improvement in system performance, the computational complexity increased considerably.
On the other hand, however, none of above studies have successfully addressed power allocation problems with continuous action space.
These methods have to set several optional power levels within the given power range for agent to choose from, and thus the final scheme is usually not optimal.

Motivated by this, in this letter, %following advanced works in the field of RL \cite{DDPG, ExperienceBuffer},
we propose a prioritized experience replay aided deep deterministic policy gradient (PER-DDPG) learning framework for the outage probability minimization problem.
%The agent can jointly optimize its action policy for discrete relay selection and continuous power allocation, to minimize the outage probability in a two-hop relay network.
The proposed method performs efficient experience learning, and realizes precise control of continuous action by performing gradient operation directly on the action policy, with which the agent can optimize its action policy for discretized relay selection and continuous power allocation.
%In addition, to evaluate the performance of our proposed learning framework in extreme situations, we design an outage-based reward function.
%The reward fed back from environment is only determined by a binary signal that represents success or failure of communication, rather than other return with concrete expression such as SNR.
%The rest of this letter is organized as follows. Section \ref{sect model} analyzes the communication system model and formulates the optimization problem. Section \ref{sect solution} gives the detail of our proposed method. Section \ref{sect Result} presents the simulation results. Finally, Section \ref{sect Conclusion} concludes the letter with a summary.

\section{System Model and Problem Formulation}\label{sect model}
\begin{figure}[ht]
	\centering
	\includegraphics[scale=0.58]{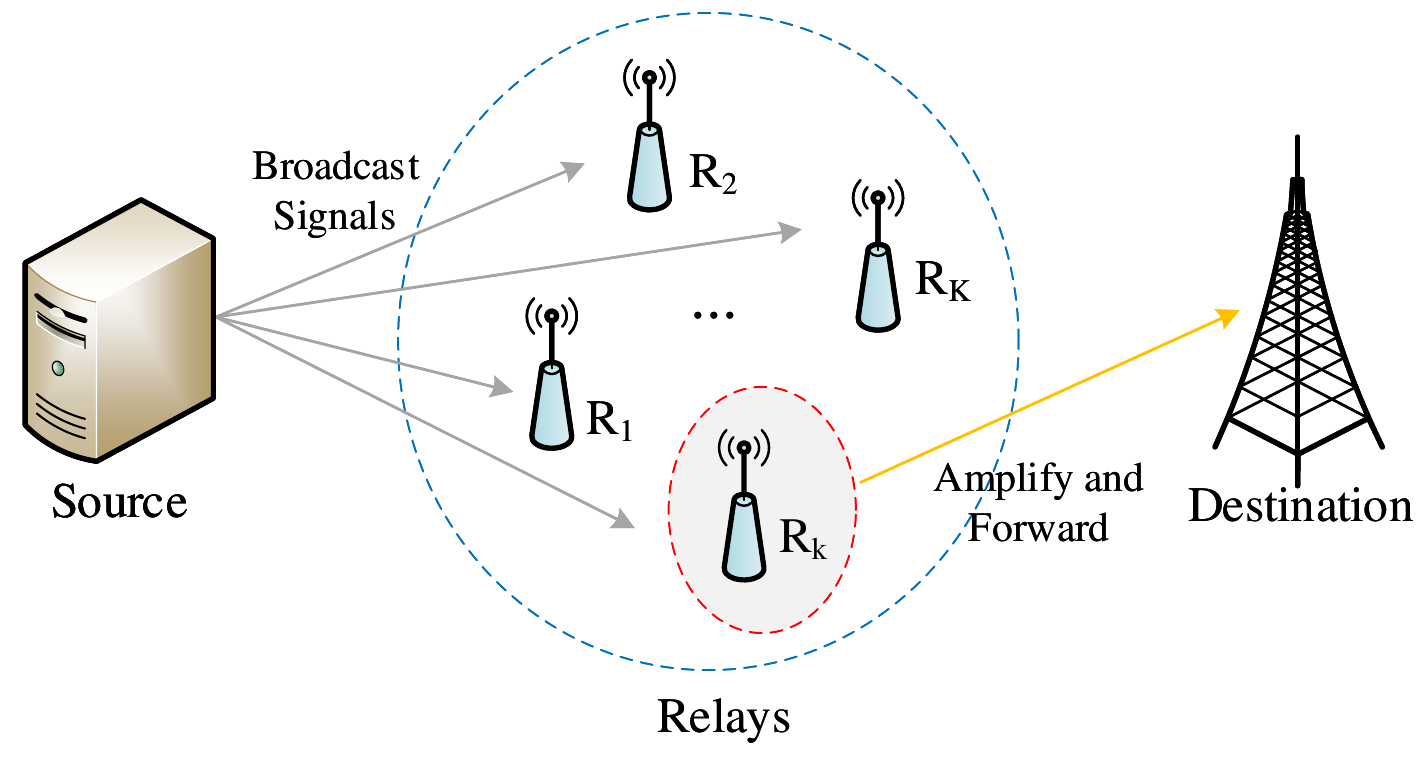}
    \caption{Cooperative relay network.}
	\label{model}
\end{figure}

As shown in Fig. \ref{model}, there is an $N_S$-antenna source $S$, an $N_D$-antenna destination $D$, and a group of single-antenna relays $R=\{R_1,R_2,\dots,R_K\}$ in the two-hop wireless relay network.
Suppose the source is far from the destination, and it does not have the direct link to destination.
Therefore, the relay which uses amplify-and-forward (AF) protocol to process the received signal, is needed to help communication.

We consider a half-duplex signaling mode because of equipment limitation. %, and only one relay is selected.
The source selects only one relay and orthogonal channels are used, in order to achieve full set gain and avoid mutual interference.
Therefore, the communication from source $S$ to destination $D$ via the selected relay $R_k$ will take two time slots.

In the first time slot, the source broadcasts its signal, then all candidate relays listen to this transmission.
The received signal at $R_k$ can be written as
\begin{equation}\label{eq3_1} y_{sk}(t)=\sqrt{P_s}\boldsymbol w_{s}^{\dagger} \boldsymbol h_{sk}(t)x(t)+n_k(t), \end{equation}
where $P_s \in [0,P_{max}]$ represents the transmission power at source, $x(t)$ represents data symbol.
$\boldsymbol h_{sk}$ represents channel vector between source and relay $R_k$, where each element is a complex Gaussian random variable with zero mean and variance $\sigma_{sk}^2$, and $n_k$ is the complex Gaussian noise with variance $\sigma_n^2$ at relay.
$\boldsymbol w_{s}^{\dagger}=\boldsymbol h_{sk}/\|\boldsymbol h_{sk}\|$ is the normalized beamforming vector using principles of maximal ratio transmission, where $\dagger$ denotes conjugate transpose operation.

In the second time slot, the selected relay amplifies and forwards the detected signal to destination.
Then the received signal at destination can be written as
\begin{equation}\label{eq3_3} \begin{aligned}
\boldsymbol y_{kd}(t)=\sqrt{P_r}\boldsymbol h_{kd}(t)\beta y_{sk}(t)+\boldsymbol n_d(t),
\end{aligned}\end{equation}
where $P_r\in [0,P_{max}]$ is the transmission power at relay, and similarly, $\boldsymbol h_{sk}$ represents channel vector between relay and destination, $\boldsymbol n_d\sim \mathcal{CN}(\boldsymbol 0,\sigma_n^2\boldsymbol I_{N_D})$ is the complex Gaussian noise at destination.
By employing maximal ratio combining methods, we multiply signal $\boldsymbol y_{kd}$ by a beamforming vector $\boldsymbol w_{d}^{\dagger}=\boldsymbol h_{kd}/\|\boldsymbol h_{kd}\|$, and have
\begin{equation}\label{eq3_3_2} \begin{aligned}
x_{kd}(t)=\sqrt{P_s P_r} &\boldsymbol w_{d}^{\dagger} \boldsymbol h_{kd}(t) \beta \boldsymbol w_{s} \boldsymbol h_{sk} x(t) +  \\
& + \sqrt{P_r} \boldsymbol w_{d}^{\dagger} \boldsymbol h_{kd}(t) \beta n_k(t) + \boldsymbol w_{d}^{\dagger} \boldsymbol n_d(t),
\end{aligned}\end{equation}
where $\beta^2 =(P_s\|\boldsymbol h_{sk}\|^2+\sigma^2_n)^{-1}$ is the amplification factor.
%\begin{equation}\label{eq3_4} \beta =\sqrt{\frac{1}{P_s\|\boldsymbol h_{sk}\|^2+\sigma^2_n}}. \end{equation}

Similar to \cite{9137340, 5710995}, we have the final end-to-end SNR $\varphi_{z} =\varphi_{sk}\varphi_{kd} / (\varphi_{sk}+\varphi_{kd}+1)$ after some manipulations,
%\begin{equation}\label{eq3_5} \varphi_{z} =\frac{\varphi_{sk}\varphi_{kd}}{\varphi_{sk}+\varphi_{kd}+1}, \end{equation}
where $\varphi_{sk}=P_s{\|\boldsymbol h_{sk}\|}^2/\sigma_n^2$ and $\varphi_{kd}=P_r{\|\boldsymbol h_{kd}\|}^2/\sigma_n^2$.
Then we have the mutual information (MI) between source and destination using unit bandwidth.
\begin{equation}\label{eq3_7} \begin{aligned}
I=\frac{1}{2}\log_2(1+\frac{\varphi_{si}\varphi_{id}}{\varphi_{si}+\varphi_{id}+1}).
\end{aligned}\end{equation}

We assume that the RL agent has access to CSI in the previous time slot, which can be denoted as $\boldsymbol h(t)=\{\boldsymbol h_{sk}(t-1),\boldsymbol h_{kd}(t-1)\}$.
Then we define the following indicator function to represent the outage event.
\begin{equation}\label{eqpro1} f(t) \triangleq f\big(R_k(t),P_s(t),P_r(t);\boldsymbol h(t)\big) \triangleq \mathbbm{1}_{I<\lambda}, \end{equation}
where $\lambda>0$ denotes the outage threshold, and $\mathbbm{1}_{I<\lambda}$ denotes the indicator function which equals to 1 when the inequality $I<\lambda$ is satisfied.
%Consider the fact that
%\begin{equation}\label{face}
%\lim\limits_{T\rightarrow\infty} F(R_k,P_s,P_r; \boldsymbol h) \triangleq \frac{1}{T}\sum\limits_{t=1}^T f(t),
%\end{equation}
Since the expectation of an indicator function can be used to calculate the probability of its original event, we then formulate the optimization problem for minimizing outage probability as follows.
\begin{equation}\label{eqpro2} \begin{aligned}
&\boldsymbol P_1: \min\limits_{R_k(t),P_s(t),P_r(t)} \mathbb{E}\Bigg[\frac{1}{T}\sum\limits_{t=1}^T f(t)\Bigg] \\
\textit{s.t.}\  &\textbf C_1: R_k \in \{R_1,R_2,\dots,R_K\}, \\
&\textbf C_2: 0 \leq P_s,P_r \leq P_{max}, \\
&\textbf C_3: P_s+P_r \leq P_{max},
\end{aligned}\end{equation}
where $\mathbb{E}[\cdot]$ denotes expectation operation,
and our goal is to minimize the average outage probability over $T$ time slots.

\section{Deep Reinforcement Learning Method}\label{sect solution}
Without acquiring real-time CSI, we turn to RL methods for solutions.
In our method, the source node acts as RL agent, and selects a proper relay and sets transmission power according to its observation of historical CSI, then it receives communication result as reward from the environment. %, which indicates whether the communication is successful.
In this section, we model this process as a Markov decision process (MDP), and then describe our proposed method.
\subsection{Markov Decision Process and System Variables}\label{subsect MDP}
The MDP consists of environment $\mathcal{E}$, state space $\mathcal{S}$, action space $\mathcal{A}$, and reward space $\mathcal{R}$. At each time step $t$, RL agent observes current state $s_t\in\mathcal{S}_t$, and accordingly selects action $a_t\in\mathcal{A}_t$. After executing action $a_t$, it receives a scalar reward $r_t \in \mathcal{R}_t$ from the environment $\mathcal{E}$ and observers next state $s_{t+1}$. %according to an unknown transition probability $p(s_{t+1}|s_t,a_t)$.
This process will continue until terminal state is reached.
Specifically, components of our system are designed as follows.

\begin{itemize}
    \item \textbf {Environment:}
    The environment is a virtual scenario which corresponds to the two-hop cooperative communication system established in section \ref{sect model}.

    \item \textbf {RL Agent:}
    In our proposed method, the source node is equipped with learning ability, and is considered as RL agent.
    Note that, any information about the environment is unknown to RL agent at the beginning.
    Therefore, RL agent needs to act and interact with the environment to obtain experiences for strategy learning.

    \item \textbf {System State:}
    Full observation of environment consists of channel states between any two nodes in the previous time slot. Therefore, we consider historical channel state $\boldsymbol h(t)$ as system state, which can be denoted as
    \begin{equation}\label{eqstate_space} \mathcal{S}_t \triangleq [\boldsymbol h_{sk}(t-1),\boldsymbol h_{kd}(t-1)], \end{equation}
    where $\boldsymbol h_{sk}(t-1)$ represents the set of channel vectors between source and all relays in time slot $t-1$, and similarly $\boldsymbol h_{kd}(t-1)$ represents that between all relays and destination.
    %In addition, in our MDP model, channel states of adjacent time slots is set to change according to a transition probability $p(s_{t+1}|s_t,a_t)$, which the agent has no prior knowledge about.
    More specifically, in our simulation environment, channel states of adjacent time slots will change according to the following Gaussian Markov block fading autoregressive model \cite{arXiv:1812.07394, 5710995}, which the RL agent has no prior knowledge about.
    \begin{equation}\label{channel_change} \boldsymbol h_{ij}(t)=\rho\boldsymbol h_{ij}(t-1)+\sqrt{1-\rho^2}\boldsymbol e(t),  \end{equation}
    where $\rho$ denotes the normalized channel correlation coefficient, and $\boldsymbol e(t)\sim\mathcal{CN}(\boldsymbol 0,\sigma^2\boldsymbol I)$ denotes the error variable and is uncorrelated with $\boldsymbol h_{ij}(t)$.

    \item \textbf {System Action:}
    In each time slot, RL agent needs to select relay and make power allocation simultaneously. Therefore, our system action can be defined as
    \begin{equation}\label{eqaction space} \mathcal{A}_t \triangleq [a^{R}(t), a^{P_s}(t)], \end{equation}
    where $a^{R}(t) \in \{1,2,\dots,K\}$ and $a^{P_s}(t) \in [0, P_{max}]$.
    Note that, the action for $P_r(t)$ is omitted, for it can be replace by the subtraction of $P_{max}$ and $P_s(t)$.

	\item \textbf {Reward Function:}
    %Reward is given from the communication environment to evaluate the executed action.
    In this letter, we consider an extreme case that the only feedback our agent gets is a binary result of successful or unsuccessful communication, which can be denoted as $\mathcal{R}_t=\{0,1\}$.
    Accordingly, we design the following outage-based binary reward function.
    %It only uses binary signals representing success or failure in communication, which can be denoted as $\mathcal{R}=\{0,1\}$.
    %According to (\ref{eqpro1}), the reward received at time slot $t$ will be calculated by the following reward function.
    \begin{equation}\label{eqreward}
    r_t = r(s_t,a_t) \triangleq 1-f\big(a^{R}(t), a^{P_s}(t);\boldsymbol h(t)\big).
    \end{equation}
\end{itemize}

The total accumulated reward at time step $t$ can be written as $R_t = \sum\nolimits_{i=t}^T\gamma^{i-t}r_i$, where $T$ denotes total step and $\gamma\in[0,1]$ denotes discount factor.
The goal of RL agent is to maximize the expected accumulated reward from each state $s_t$.

\subsection{PER-DDPG Solution}\label{subsect DDPG}
To achieve optimal actions under different states, the action-value function $Q(s_t,a_t;\theta)=\mathbb{E}[R_t|s_t,a_t]$ is first defined to describe the expected return after selecting action $a_t$ in state $s_t$, which is controlled by network parameter $\theta$.
Then, the optimal action-value function is denoted as $Q^{\ast}(s_t,a_t;\theta)=\max_{a_t\in\mathcal{A}_t}Q(s_t,a_t;\theta)$, which obeys Bellman function.
\begin{equation}\label{eqvalueoptimal}
Q^{\ast}(s_t,a_t;\theta)=\mathbb{E}_{s_{t+1}\sim\mathcal{E}}[r+\gamma\max\limits_{a_{t+1}}Q(s_{t+1},a_{t+1};\theta)],
\end{equation}

\begin{figure}[h]
	\centering
    \includegraphics[scale=0.62]{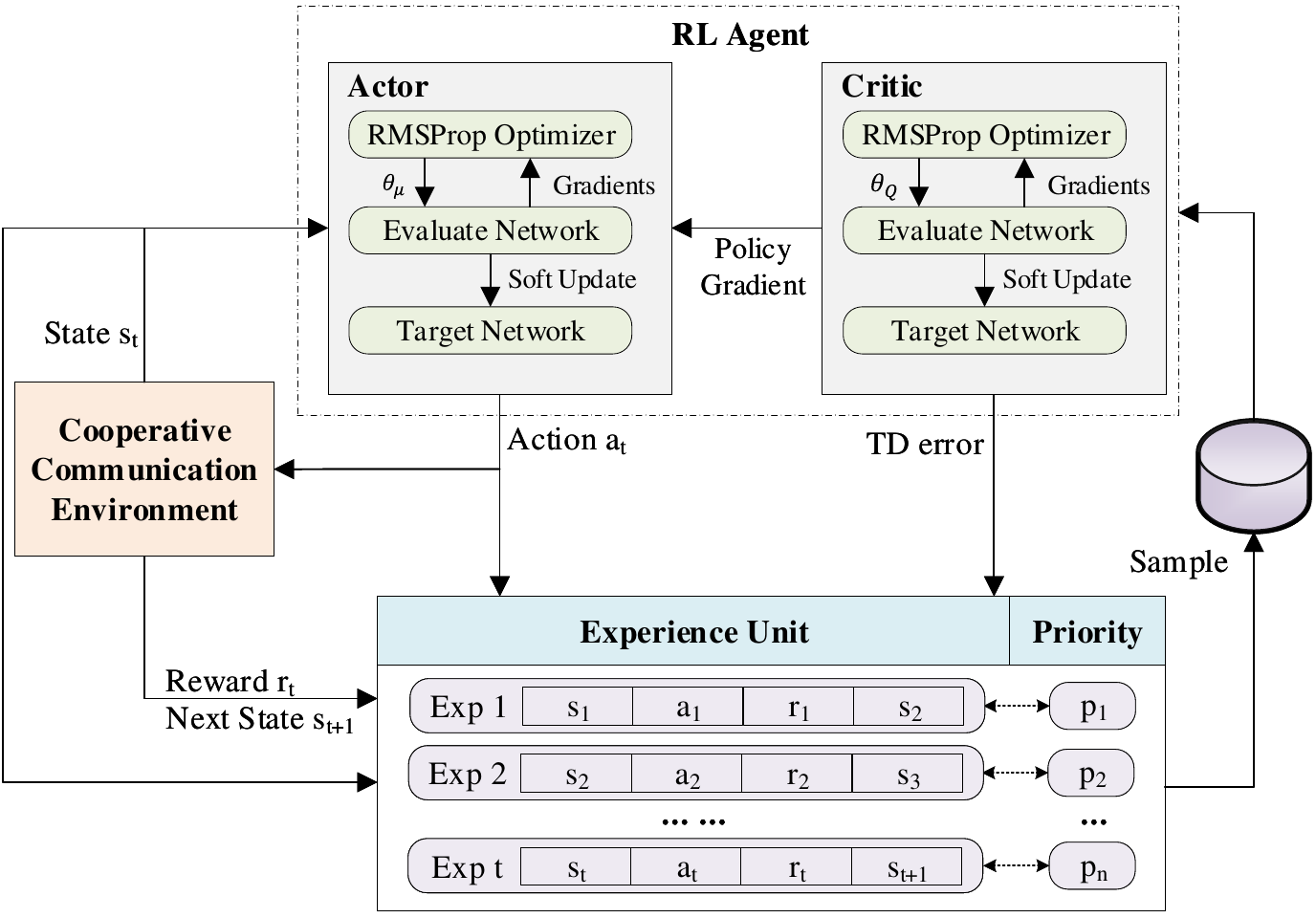}
	\caption{PER-DDPG learning framework.}
	\label{framwork}
\end{figure}

The utilization of DNN has guaranteed the ability of generalization in terms of action-value estimation.
Further, to data-efficiently deal with continuous power action space, we propose a PER-DDPG learning method, whose framework is shown in Fig. \ref{framwork}.

For sampling efficiently, we maintain a prioritized experience buffer $\mathcal{B}$, to store agent's experience unit $e_t=\{s_t,a_t,r_t,s_{t+1}\}$ after each interaction. The priority $p_t$ indicates the importance of corresponding unit, and it is intuitive to employ TD error $\delta_t$ as a proxy for priority, which specifically shows the difference between current state-action value and its next-step bootstrap estimation. The larger the value, the more RL agent can learn from this experience unit.

When calculating the priority, a small positive constant $\epsilon$ is introduced, \textit{i.e.} $p_t=|\delta_t|+\epsilon$, to ensure that each unit has the probability to be sampled even if its TD error is zero. In addition, to correct the bias illustrated in \cite{ExperienceBuffer}, we accordingly employ the following importance-sampling weight
\begin{equation}\label{ISweight}
w_i = \frac{(B_{size} \cdot p_i)^{-\kappa}}{\max\nolimits_{j<t} w_j},
\end{equation}
where $B_{size}$ denotes the size of experience buffer, and $\kappa$ is an exponent between 0 and 1.

In terms of RL agent, it consists of two parts, which are called actor and critic, respectively \cite{DDPG}. Further, the agent employs two separate DNNs for both of them, which are known as evaluate network and target network.

\textbf{Critic: }
The critic estimates action-value function by employing a DNN with parameter $\theta_Q$.
During training process, the agent samples a mini-batch of experience units according to experience priority.
Similar to other value-based RL methods, the critic tries to minimize the following loss function.
\begin{equation}\label{eqcriticloss1}
L(\theta_Q)=\mathbb{E}_{e_t\sim\mathcal{B}}\Big[w_t \cdot \delta_t^2(s_t, a_t; \theta_Q)\Big]
\end{equation}
with TD error represented as
\begin{equation}\label{eqcriticloss2}
\delta_t(s_t, a_t; \theta_Q) = r_t + \gamma \max\limits_{a_{t+1}}Q(s_{t+1}, a_{t+1}; \theta^{-}_Q) - Q(s_t, a_t; \theta_Q),
\end{equation}
where $\theta^{-}_Q$ is a group of old parameters in target network.
In order to improve learning stability, old parameters will be soft replaced periodically following $\theta^{-}_Q \leftarrow \tau\theta_Q + (1-\tau)\theta^{-}_Q$ with $\tau \ll 1$.
Afterwards, parameters in evaluate network will be updated using RMSProp optimization.
\begin{equation}\label{eqcriticupdate}
\theta_Q \leftarrow \theta_Q - \eta_Q w_t \delta_t(s_t; \theta_Q) \nabla_{\theta_Q}Q(s_t, a_t; \theta_Q),
\end{equation}
where $\eta_Q$ denotes learning step size for critic.

\textbf{Actor: }
The actor is used to learn action policy and perform primitive actions.
It maintains a parameterized function $\mu(s;\theta_\mu)$, which specifies the current policy by deterministically mapping states to specific actions.
We define the following performance objective for current action policy.
\begin{equation}\label{eqactorperformance}
J(\theta_\mu)=\mathbb{E}_{s_t\sim\mathcal{B}}[Q(s_t,\mu(s_t;\theta_\mu);\theta_Q)].
\end{equation}

The gradient of such deterministic policy moves the action policy in the direction of the gradient of action-value function, that is, network parameters $\theta_\mu$ will be updated in proportion to $\nabla_{\theta_\mu}Q(s_t,\mu(s_t;\theta_\mu);\theta_Q)$.
By using chain rule, it can be divided into two parts as shown below.
%Following \cite{DPG}, we can derive the policy gradient of actor network.
\begin{equation}\label{eqactorgradient}\begin{aligned}
	\nabla_{\theta_\mu}J=\mathbb{E}_{s_t}&\Big[\nabla_{\theta_\mu}\mu(s_t;\theta_\mu) \nabla_{a}Q(s_t,a_t;\theta_Q)|_{a_t=\mu(s_t;\theta_\mu)}\Big],
\end{aligned}\end{equation}
based on which the actor parameters $\theta_\mu$ is updated as follows.
\begin{equation}\label{eqactorupdate}
\theta_\mu \leftarrow \theta_\mu - \eta_\mu\nabla_{\theta_\mu}J,
\end{equation}
where similarly $\eta_\mu$ denotes learning step size for actor.

Pseudocode of our method can be found in Algorithm \ref{algoDDPG}.
\begin{algorithm}[htb]
	\caption{PER-DDPG for Relay and Power Optimization}
	\label{algoDDPG}
	\begin{algorithmic}[1]
        \STATE Initialize experience buffer $\mathcal{B}=\varnothing$.
		\STATE Initialize evaluate network parameters $\theta_Q$ and $\theta_\mu$.
		\STATE Initialize target network with $\theta^{-}_Q=\theta_Q$ and $\theta^{-}_\mu=\theta_\mu$.
		
		\FOR{episode $u=1,2,\dots,u_{max}$}
		\STATE Initialize communication environment, get state $s_1$.
        \STATE Initialize a random process $\Delta\mu$ as noise.
		\FOR{time slot $t=1,2,\dots,t_{max}$}
		\STATE Choose action $a_t=\mu(s_t;\theta_\mu)+\Delta\mu_t$ to determine the selected relay and power for transmission.
		\STATE Execute action $a_t$, then receive reward $r_{t}$ and observe next state $s_{t+1}$.
		\STATE Collect and save current experience $e_t$ with initial priority $p_t=\max_{i<t}p_i$, and then sample a mini-batch of experience units $e_j$ according to probability $p(j)=p_j^{\alpha}/{\sum\nolimits_{i}p_i^{\alpha}}$ with predefined exponent $\alpha$.
		\STATE Calculate importance-sampling weight and TD error of each experience unit according to (\ref{ISweight}) and (\ref{eqcriticloss2}).
        \STATE Minimize the loss of mini-batch in (\ref{eqcriticloss1}), and update evaluate network of critic according to (\ref{eqcriticupdate}).
		\STATE Calculate the sampled policy gradient in (\ref{eqactorgradient}), and update evaluate network of actor according to (\ref{eqactorupdate}).
		\STATE Update experience priority using $p_j=|\delta_j|+\epsilon$.
        \STATE Update parameters of corresponding target networks by $\theta^{-}_Q \leftarrow \tau\theta_Q + (1-\tau)\theta^{-}_Q$  and $\theta^{-}_\mu \leftarrow \tau\theta_\mu + (1-\tau)\theta^{-}_\mu$.
		\ENDFOR
		\ENDFOR
	\end{algorithmic}
\end{algorithm}

\section{Evaluation}\label{sect Result}
In this section, we present implementation details of simulation environment, and demonstrate performance of the proposed method.
Similar to the settings in \cite{9137340}, the required outage threshold is set as $\lambda=0.1$,
%The number of candidate relay nodes $K$ is 20,
and total maximum power $P_{max}$ for source and relay transmission is 1W.
Learning rates for updating critic network and actor network are set as $\eta_Q = 0.005$ and $\eta_\mu = 0.001$, respectively. Parameter for soft update is set as $\tau=0.001$.
In addition, the size of experience buffer $B_{size}$ is 10000, and mini-batch size which determines numbers of training cases is 128.

Given the lack of knowledge of the underlying channel distribution in actual communication system, we mainly employ random selection and existing RL algorithms as baseline methods.
More specifically, since the PER operation can be decoupled from our proposed method, we are able to take original DDPG scheme for comparison.
In addition, DQN is a widely used RL method in the field of communication, which we will also take as a baseline method.
Note that, DQN can only solve problems with discrete action spaces.
Therefore, when using this method, we divide the power into $L$ power levels for the agent to choose from, which can be denoted as $\frac{1}{L}P_{max},\frac{2}{L}P_{max},\dots,P_{max}$.

We first evaluate the training performance of different methods with total training episodes $\mu_{max}=100$.
Specifically, we first perform 10 training trials for each method.
Then, we select the successful trials among these 10 repetitions, and use solid line to represent the median and shadow region to represent the range of them.
%The training performance and statistical analysis are depicted in Fig. \ref{TrainReward} and Table \ref{training_result}, respectively.

\begin{figure}[ht]
	\centering
    \includegraphics[scale=0.55]{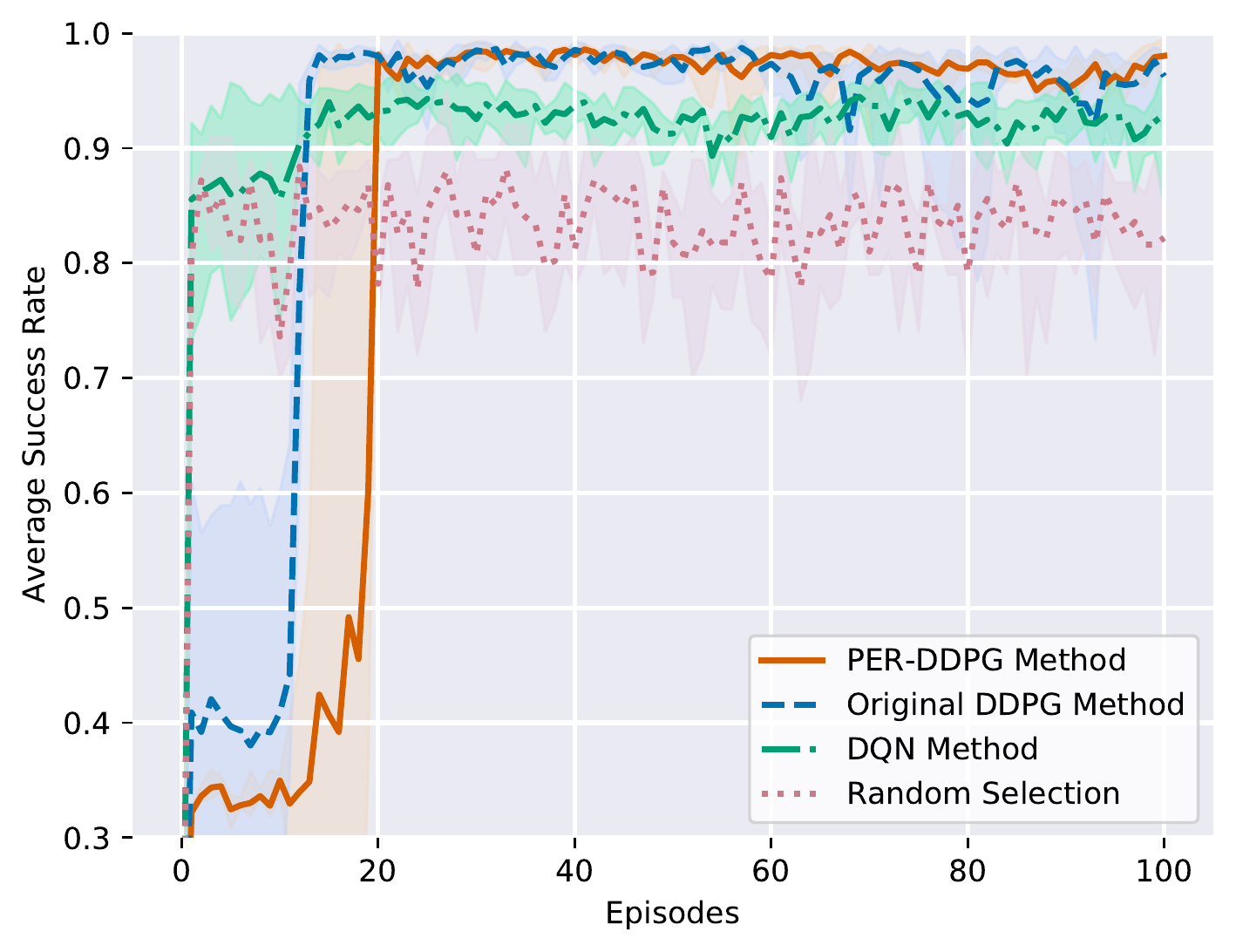}
	\caption{Average success rate with different methods.}
	\label{TrainReward}
\end{figure}

As shown in Fig. \ref{TrainReward}, the training performance of random selection is poor, while all DRL methods can converge.
Before training with our proposed method (orange line) and its another version without handling experience replay (blue line), we both use 10 episodes for RL agent to select randomly and collect enough experience units.
We find that, PER-DDPG and original DDPG can finally converge to a similar value.
%We find that our PER-DDPG can converge faster with slighter fluctuation, which reflects the efficiency in data sampling and learning.
When compared with DQN method (green line), our method can achieve a better result with an improvement of about $4\%$ in average success rate.
%In addition, when the number of optional power levels increases from 10 to 100, the average success rate by using DQN does not improve as expected, but decreases by about $10\%$.
%It vividly shows that DQN method can not deal with high-dimensional or even continues action space, and too many optional actions will make RL agent fails to learn proper action policy, while our method succeeds.

For better explanation, we also present the statistic for all successful training trials in Table \ref{training_result}, where we analyze the performance of the last 40 episodes (\textit{i.e.} Episode 61-100, after training curves converge).
    \begin{table}[h]
    	\centering
        \caption{Statistic of 10 trials with different methods.}
		\label{training_result}
		\begin{tabular}{|c|c|c|c|}
        \hline
			Method & Successful trials & Mean & Standard deviation\\
            \hline
			PER-DDPG & 10 & 0.969 & $1.374*10^{-2}$ \\
			Original DDPG & 9 & 0.955 & $3.209*10^{-2}$ \\
			DQN & 10 & 0.926 & $2.003*10^{-2}$ \\
            Random & - & 0.835 & - \\
        \hline
		\end{tabular}
	\end{table}

We find that with our PER-DDPG method, RL agent can obtain appropriate action policy every time, but there is one record of failure using its original version.
What's more, we observe that although original DDPG trains faster, our PER-DDPG method has smaller standard variance.
As vividly depicted in Fig. \ref{TrainReward}, the fluctuation of PER-DDPG's training curve is slighter, and the shadow region is smaller.
Actually, such failure in using original DDPG is caused by the interplay between the actor and critic updates \cite{ApproximationError}.
The usage of outdated experience unit can lead to a high variance in value estimate, which then misleads policy updates.
However, with the help of PER, we realize efficient data sampling, and can achieve better stability during the training process.

\begin{figure}[ht]
	\centering
    \includegraphics[scale=0.55]{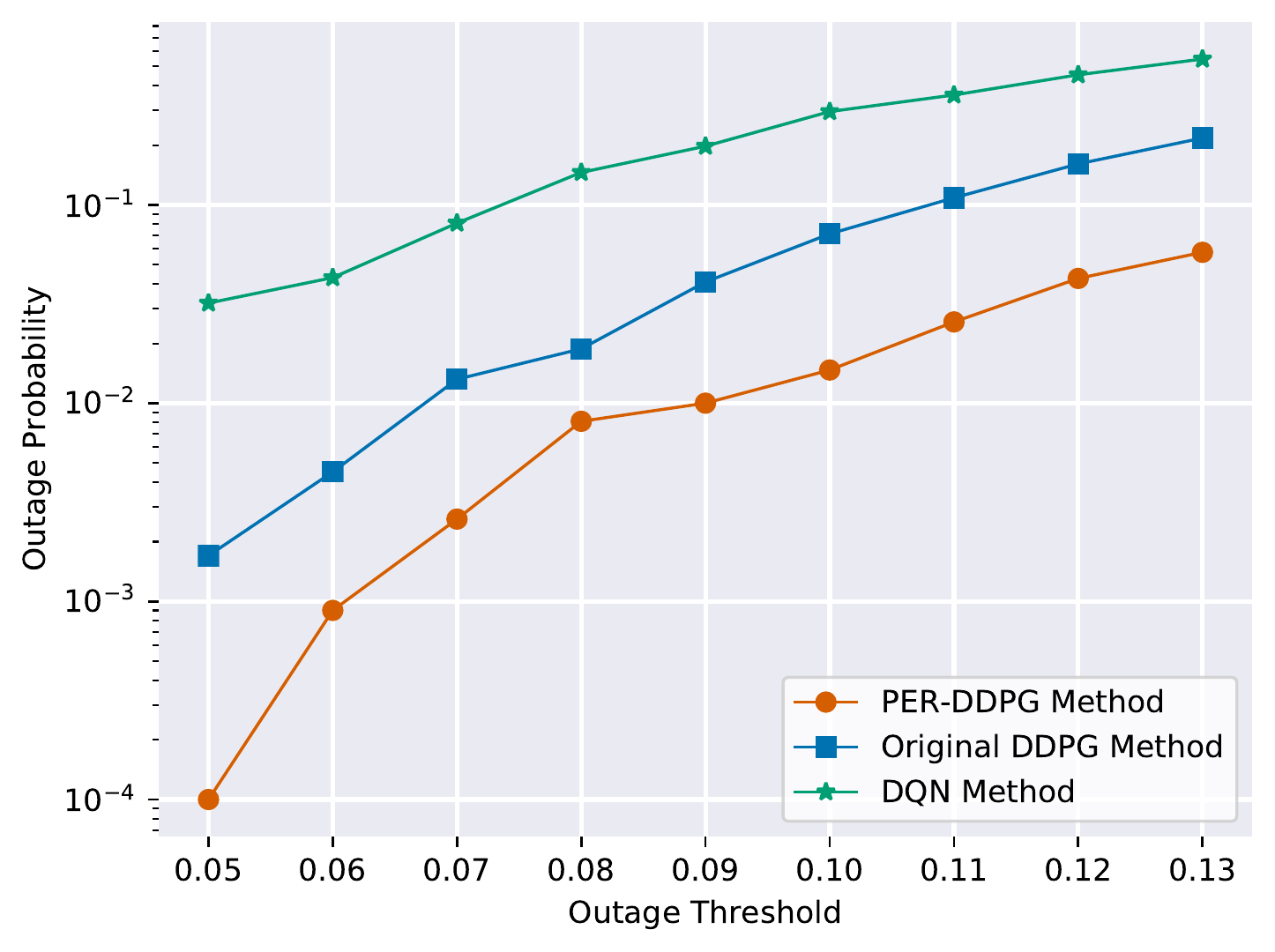}
	\caption{Testing result under different outage thresholds.}
	\label{TestReward}
\end{figure}

In testing process, we evaluate the outage probability of these well-trained DRL models with different threshold requirements.
Note that, corresponding to the training process, we preserve all successful trained network models for each method.
Then we record the average performance of each method.
The testing result is depicted in Fig. \ref{TestReward}, where we can obviously find that our PER-DDPG model still has better performance than other models.
Although original DDPG method can achieve an average success rate close to that of PER-DDPG method in training stage, due to its larger variance, the results obtained in test process are not good.
%The original DDPG model which is not equipped with prioritized experience replay performs well only when outage threshold is easy to meet, but the result deteriorates rapidly under harsh conditions.
On the other hand, DQN model always performs worse than model with policy gradient methods, for it can only perform coarse-grained discrete power optimization.
%On the other hand, DQN model trained with a larger action space always performs worse than that with a smaller action space.
%We can find the reason from Fig. \ref{TrainReward}, the fluctuation after convergence of DQN method with $L=100$ is obviously larger than the others.
%It indicates that its action policy is not robust enough to deal with the current situation, let alone be further applied to other situations.
From the above, our proposed method can obtain a robust action policy, which can effectively reduce outage probability and be applied to other situations.

\section{Conclusion}\label{sect Conclusion}
In this letter, we introduce deep deterministic policy gradient into dynamic relay selection and power optimization in a two-hop cooperative relay network, and realize data efficiency with the help of prioritized experience. %in order to minimize outage probability under a total transmission power constraint.
Unlike traditional studies, our method does not rely on any assumptions about channel distribution, and can effectively deal with optimization variables which have continuous action space.
Simulation results reveal that with our PER-DDPG method, the average communication success rate can be increased by about $4\%$ compared to existing RL methods.

\bibliographystyle{IEEEtran}
\bibliography{paper_ref}

\end{document}